\documentclass[twocolumn,superscriptaddress,showpacs,preprintnumbers,amsmath,amssymb]{revtex4}
\usepackage{graphicx}
\usepackage{dcolumn,xcolor,ulem}
\usepackage{amsmath} 
\usepackage{amssymb}
\usepackage{bm}
\usepackage[latin1]{inputenc}
\usepackage{mathrsfs}

\newcommand{\D}{\mathscr{D}}

\begin{document}

\widowpenalty=1000
\clubpenalty=1000

\preprint{APS/123-QED}

\title{Elastic turbulence in shear banding wormlike micelles\\}

\author{M.A.~Fardin}
\affiliation{Laboratoire Mati\`ere et Syst\`emes Complexes, CNRS UMR 7057\\ Universit\'e Paris Diderot, 10 rue Alice Domon et L\'eonie Duquet, 75205 Paris C\'edex 13, France} 
\affiliation{Department of Mechanical Engineering\\ Massachusetts Institute of Technology, 77 Massachusetts Avenue, MA 02139-4307 Cambridge, USA}
\author{D.~Lopez}
\author{J.~Croso}
\author{G.~Gr\'egoire}
\author{O.~Cardoso}
\affiliation{Laboratoire Mati\`ere et Syst\`emes Complexes, CNRS UMR 7057\\ Universit\'e Paris Diderot, 10 rue Alice Domon et L\'eonie Duquet, 75205 Paris C\'edex 13, France} 
\author{G.H.~McKinley}
\affiliation{Department of Mechanical Engineering\\ Massachusetts Institute of Technology, 77 Massachusetts Avenue, MA 02139-4307 Cambridge, USA}
\author{S.~Lerouge}
\altaffiliation[Corresponding author ]{}
\email{sandra.lerouge@univ-paris-diderot.fr}
\affiliation{Laboratoire Mati\`ere et Syst\`emes Complexes, CNRS UMR 7057\\ Universit\'e Paris Diderot, 10 rue Alice Domon et L\'eonie Duquet, 75205 Paris C\'edex 13, France}

\date{\today}

\begin{abstract}
We study the dynamics of the Taylor-Couette flow of shear banding wormlike micelles. We focus on the high shear rate branch of the flow curve and show that for sufficiently high Weissenberg numbers, this branch becomes unstable. This instability is strongly sub-critical and is associated with a shear stress jump. We find that this increase of the flow resistance is related to the nucleation of turbulence. The flow pattern shows similarities with the elastic turbulence, so far only observed for polymer solutions. The unstable character of this branch led us to propose a scenario that could account for the recent observations of Taylor-like vortices during the shear banding flow of wormlike micelles.\\
\end{abstract}

\maketitle

Wormlike micelles are elongated, polymer-like structures resulting from the self-assembly of amphiphilic molecules in aqueous solution~\cite{Berret05,Cates90}. In contrast to regular polymers, they continuously break and fuse, providing an additional relaxation mechanism. In the fast breaking regime, Cates' reptation-reaction model~\cite{Cates87} predicts that wormlike micelles solutions relax mono-exponentially with a single time $\tau_R\sim \sqrt{\tau_b\tau_r}$, where $\tau_b$ and $\tau_r$ are the breaking and reptation times.\\
In addition to their structural analogy, polymers and wormlike micellar solutions can exhibit flow instabilities when submitted to even moderate shear rates. In particular, many wormlike micelles solutions have been observed to undergo a shear banding transition. Under simple shear, the base scenario is the following~\cite{Berret05,Salmon2003}~: below a critical shear rate $\dot\gamma_l \sim 1/\tau_R$, the flow is homogeneous. Above $\dot\gamma_l$, the system becomes mechanically unstable. A phase of lower viscosity nucleates, inducing a banded state in which the initial viscous phase and the fluid phase coexist at constant stress. Increasing the imposed shear rate only affects the relative proportions of each bands, up to a second critical value  $\dot\gamma_h$, where the high shear rate phase entirely fills the flow geometry. Beyond $\dot\gamma_h$, the homogeneity of the flow is recovered. Albeit extremely well documented, the shear banding instability is still not fully understood~\cite{Lerouge09,Fielding07,Cates06}. In particular, it has been shown recently in cylindrical Couette geometry that contrary to the usual view, the shear banding flow may not be purely one-dimensional, but instead is organized into Taylor-like vortices stacked along the vorticity direction. These cellular structures are mainly localized in the high shear rate band and exhibit a complex dynamics depending on the applied shear rate~\cite{Fardin09}. \\
Recent theoretical developments have tried to rationalize the 3D flow, by involving an interfacial mechanism driven by a jump in normal stresses across the interface between bands~\cite{Fielding2005,Fielding07}. Besides, complex 3D flow structures are also well-known to develop at low Reynolds number $Re$ in polymer solutions~\cite{Larson90,Larson92,Baumert97}. The underlying instabilities are a consequence of the non-Newtonian stress field that is created in the flow due to elasticity. The importance of the elastic nonlinearity is expressed by the Weissenberg number, defined as $Wi\equiv \dot\gamma\tau_R\equiv N_1/\sigma$, where $N_1$ is the first normal stress difference and $\sigma$ is the shear stress. To a good extent, those two definitions are equivalent (see Supplementary Figure). Thereafter we only use $Wi\equiv \dot\gamma\tau_R$. 
$Wi$ acts as a control parameter analogous to $Re$ in Newtonian fluids. When increasing $Wi$, a viscoelastic fluid is likely to undergo a transition from 1D flow to various coherent 3D flows, which would eventually lead to turbulence~\cite{Groisman00,Larson00,Morozov07}.
For polymer solutions flowing in curved geometries, elastic instabilities are triggered above a threshold that follows a general criterion established by Pakdel and McKinley and corresponding to values of $Wi$ ranging typically between 1 and 10~\cite{Larson90,Mckinley96}. In wormlike micelles, the onset of the shear banding regime is characterized by $Wi\simeq~1$~\cite{Berret05}. Hence, on the stress plateau and above, the Pakdel-McKinley condition is satisfied, suggesting that the elastic instability could be responsible for the 3D shear banding flow~\cite{Fardin09}.\\
In this letter, we investigate the flow behavior of the high shear rate branch ($Wi>Wi_h \equiv \dot\gamma_h \tau_R$) of a shear banding wormlike micelles system in Taylor-Couette geometry. We show that once the induced phase fills the whole sample ($Wi\gtrsim Wi_h$), the vortex structure is destroyed and the 1D structure of the flow is recovered. Nevertheless, above a critical threshold $Wi_c$, we observe the growth of another instability leading to a random flow state, that presents all the features of the elastic turbulence. This indicates that the induced phase in shear banding wormlike systems can indeed be subject to elastic instability. From this observation, we propose a scenario that points to a bulk elastic instability as the underlying mechanism for the vortex flow in the banding regime.
\begin{figure}[t]
\begin{center}
\includegraphics[width=7cm,clip]{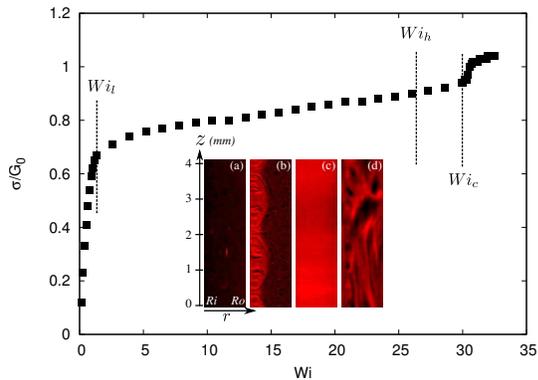}
\caption{Normalized shear stress $\sigma/G_0$ as a function of the Weissenberg number $Wi=\dot\gamma\tau_R$. The data are gathered for increasing Weissenberg numbers from 0 to 34. Inset~: Views of the velocity gradient-vorticity plane for different $Wi$. The inner and outer cylinders are respectively on the left and right sides of the pictures. (a) $Wi<Wi_l$~: homogeneous flow of the low shear rate phase. (b) $Wi_l<Wi<Wi_h$~: banded state. The interface between bands presents oscillations that scale with the size of the high shear rate band, in which vortices stacked in the vorticity direction develop. (c) $Wi_h<Wi<Wi_c$~: homogeneous flow of the high shear rate phase. (d) $Wi>Wi_c$~: turbulent flow of the high shear rate phase. 
\label{fig1}}
\end{center}
\end{figure} 

\begin{figure}[t]
\begin{center}
\includegraphics[width=7cm,clip]{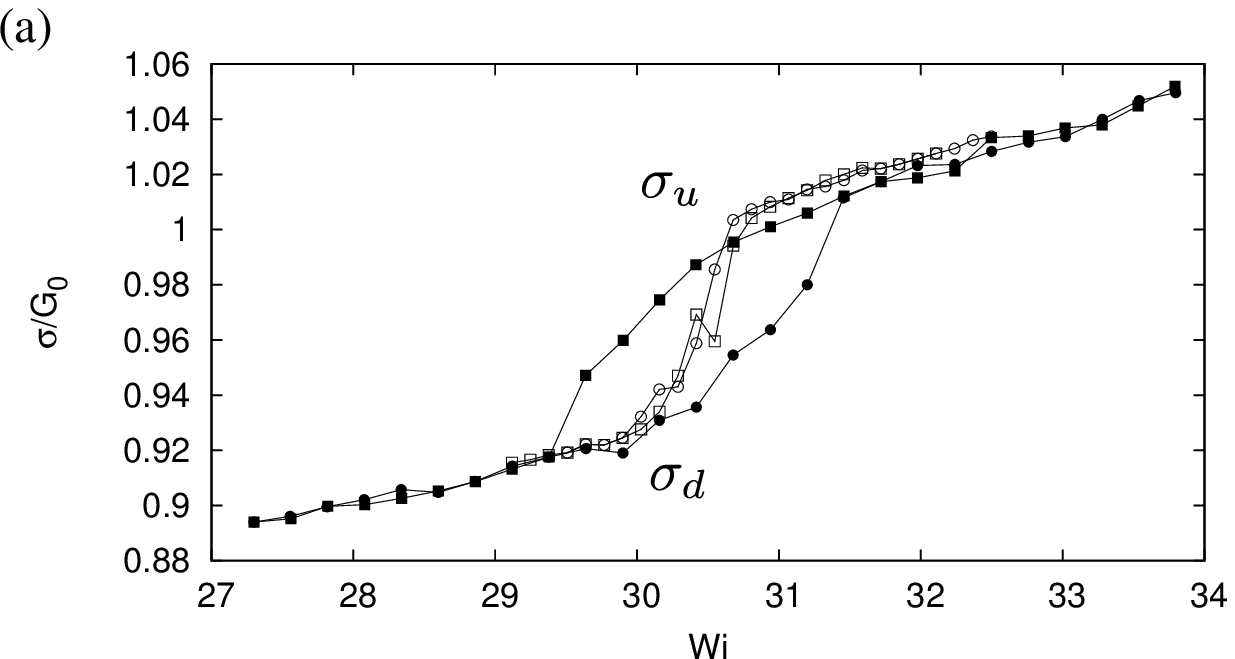}
\includegraphics[width=7cm,clip]{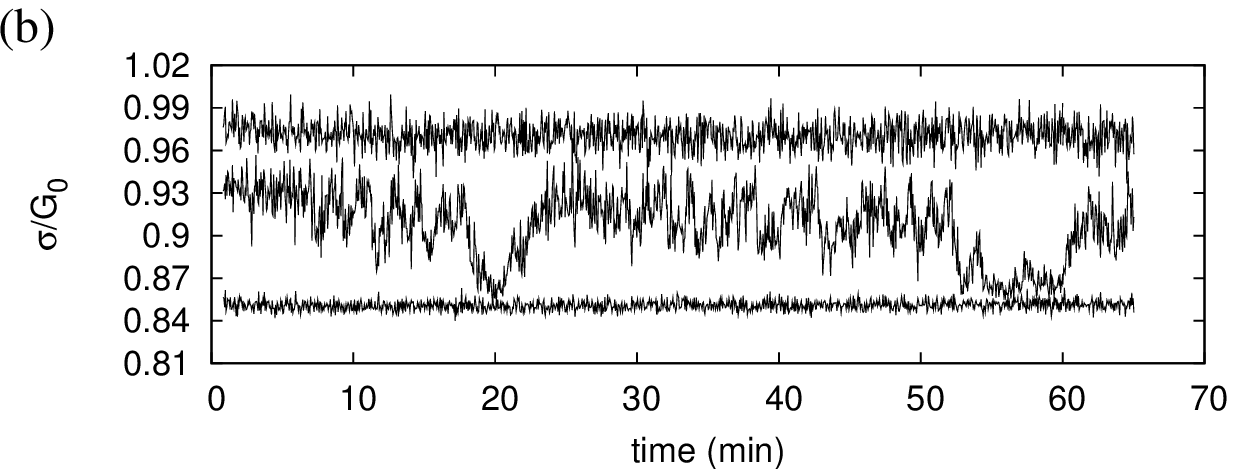}
\caption{ (a) Close-up on the high shear rate branch of the flow curve ($Wi>Wi_h$) obtained for increasing (circles) and decreasing (squares) shear rate, for two different sampling times~: 2 s/point (close symbols) and 1 min/point (open symbols). $\sigma_u$ and $\sigma_d$ are the apparent up and down boundaries of the stress jump at $Wi_c$. (b) Shear stress evolution on long time (between 1 min and 1 hour), for different $Wi$. From top to bottom: $Wi=32.8$ (fully turbulent), $Wi=31.2$ (bi-stable), $Wi=30.1$ (fully laminar).
\label{fig2}}
\end{center}
\end{figure}

\begin{figure}[t]
\begin{center}
\includegraphics[width=7cm,clip]{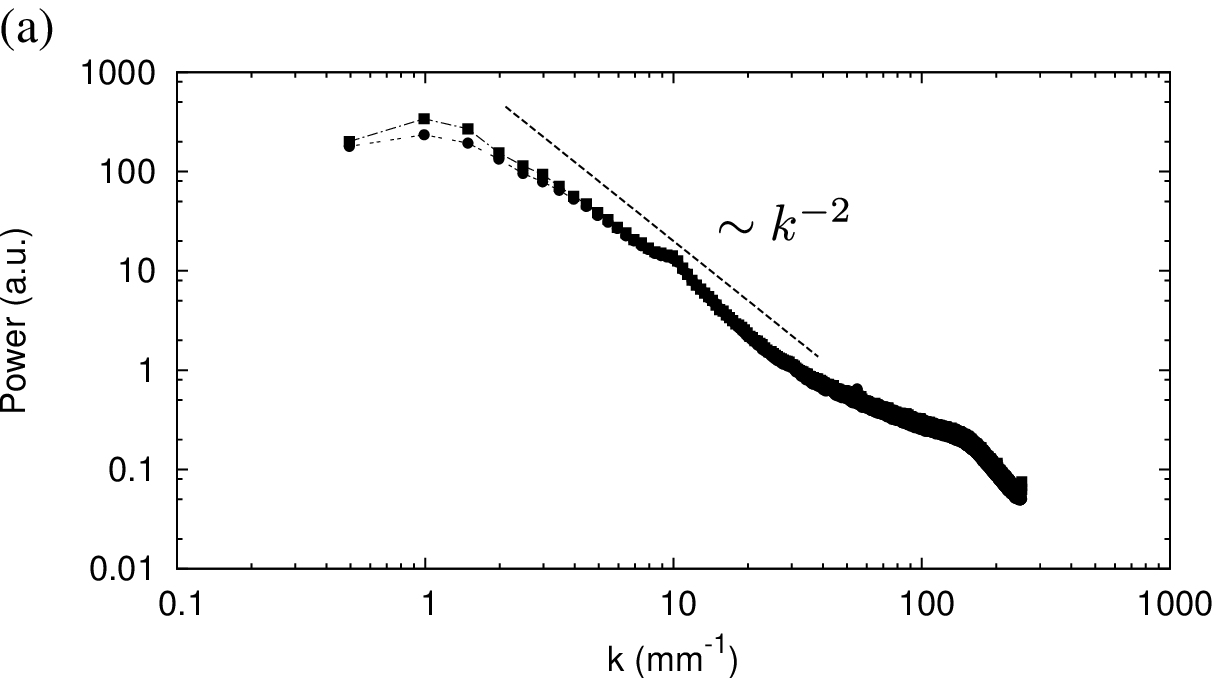}
\includegraphics[width=7cm,clip]{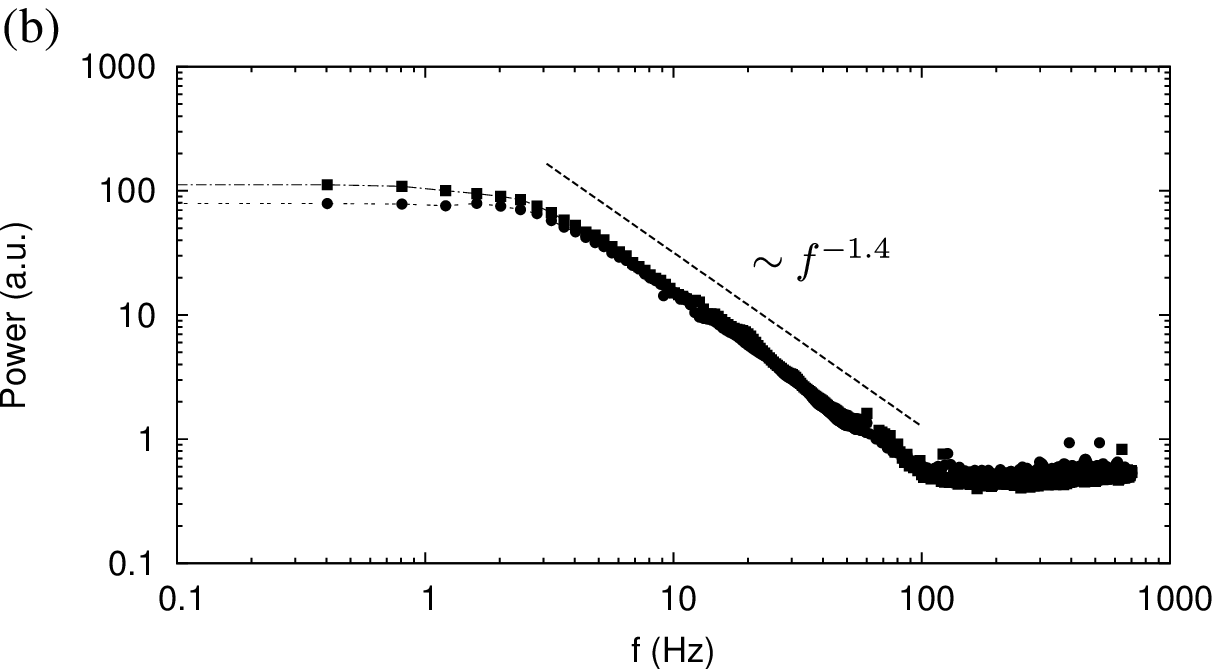}
\caption{ Average Fourier spectra of the intensity reflected by the sample in the velocity gradient-vorticity plane for two $Wi$ larger than $Wi_c$. Circles: $Wi=31.5$. Squares: $Wi=33$. (a) Spectrum obtained on the wavevector domain by computing Fourier transforms of the intensity reflected along the vertical direction at a midle point in the gap and then by averaging for different times. The dashed line is a $k^{-2}$ fit. (b) Spectrum obtained on the frequency domain by computing Fourier transforms of the intensity reflected overtime at single points in the gap and then by averaging for different points. A fast camera (Phantom V9) has been used to reach the frequency range between 10 Hz and 700 Hz. The dashed line is a $f^{-1.4}$ fit. 
\label{fig3}}
\end{center}
\end{figure}

\begin{figure}[t]
\begin{center}
\includegraphics[width=7cm,clip]{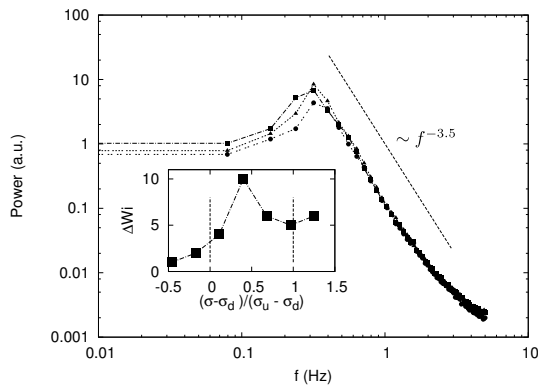}
\caption{ Average Fourier spectra of the macroscopic shear rate obtained by imposing three constant stresses in the turbulent regime. The total duration of each time serie is 5000 s and the sampling time is 0.1 s. Triangles~: $\sigma/G_0=1$. Circles~: $\sigma/G_0=1.03$. Squares~: $\sigma/G_0=1.05$. The dashed line is a $f^{-3.5}$ fit. (inset) Average amplitude of the fluctuations of shear rate (\textit{i.e. $Wi$}) for different imposed stresses before, ``during", and after the transition to turbulence. Stress is measured in unit of the stress jump in the flow curve at $Wi_c$. For stresses between $\sigma_d$ and $\sigma_u$ (\textit{i.e.} $0<\frac{\sigma-\sigma_d}{\sigma_u-\sigma_d}<1$), the system is bi-stable and exhibits large fluctuations. Spectra are computed in the fully turbulent range, for stresses larger than $\sigma_u$.
\label{fig4}}
\end{center}
\end{figure}

The sample we consider is made of cetyltrimethylammonium bromide (CTAB) at 0.3 M with sodium nitrate (NaNO$_3$) at 0.405 M. The temperature is fixed at T=28$^\circ$C. This system is Maxwellian in the linear regime, with a single relaxation time $\tau_R=0.23\pm0.02$ s and a plateau modulus $G_0=238\pm5$ Pa while, under simple shear flow, it is well-known to exhibit shear banding associated with an instability of the interface and Taylor-like vortices~\cite{Lerouge06,Lerouge08,Fardin09}. Our experiments are performed in a cylindrical Couette device with inner rotating cylinder (height $H=40$ mm, inner radius $R_i=13.33$ mm, gap $e=1.13$ mm)~\cite{Fardin09} adapted to a stress-controlled rheometer (Physica MCR301) used in strain-controlled mode. \\
Fig.~\ref{fig1} displays the overall rheological behavior of the sample for $Wi$ between 0 and 34 together with simultaneous observations in the velocity gradient-vorticity ($r,z$) plane summarizing the main flow states along the flow curve (inset). For $Wi<Wi_l$, the gap of the Couette cell appears homogeneous (inset-a) and the corresponding branch of the flow curve is slightly shear-thinning. Between $Wi_l=1.0\pm0.1$ and $Wi_h=26\pm1$, the shear stress presents a plateau and the sample splits into two shear bands of differing optical properties separated by an interface undulating along the vorticity direction. The flow is organized in Taylor-like vortices, mainly localized in the high shear rate band (inset-b)~\cite{Fardin09}. Note that the value of $Wi_h$ is given from the crossover in the flow curve between the stress plateau and the high shear branch and is in good agreement with the Weissenberg number for which the high shear rate band fills the whole gap. The standard deviation is computed from statistical measurements (see Ref.~\cite{Lerouge08} for more details). 
Let us now focus on the high shear rate branch of the flow curve  ($Wi>Wi_h$), where the induced high shear rate phase fills the entire gap. Two different regimes can be distinguished~: (1) For  $Wi$ between  $Wi_h$ and a critical value noted $Wi_c=30\pm0.5$, the flow curve is strongly shear-thinning and vortices seem to disappear (inset-c). All the methods used in ref.~\cite{Fardin09} to identify vortices in the banding regime have failed. The flow in this regime seems purely 1D. This behavior is compatible with the observations of Salmon and coworkers, who demonstrated using velocity measurements on a similar system that the induced structures were highly non-Newtonian. (2) For $Wi>Wi_c$, the flow curve presents a drastic change of variation corresponding to an apparent shear-thickening behavior. This increase of the flow resistance is due to the new structure of the flow~\cite{Groisman00}, and differs from the nucleation of a new structure in the material itself~\cite{Fischer2007}. Indeed, as seen in Fig.~\ref{fig1} (inset-d) and in the supplemental movie~\cite{movie}, in this regime, the flow appears very disorganized in space and time. This behavior is strongly reminiscent of the state of elastic turbulence highlighted by Groisman \textit{et al.} in dilute polymer solutions flowing in different flow geometries~\cite{Groisman00}. According to Ref.~\cite{Groisman00}, the main features of the elastic turbulence are the following~: fluid motion excited in a broad range of spatial and temporal scales, and significant increase of the flow resistance and rate of mixing. Furthermore, the transition to the turbulent regime is found to be hysteretic and strongly subcritical. Using mechanical measurements and direct visualizations of the sample in the ($r,z$) plane, we carefully analyse the irregular flow we observe.\\
Fig.~\ref{fig2}a shows the rheological behavior of the high shear rate branch of the flow curve during increasing and decreasing shear rate sweeps. When the shear rate is changed quickly, the transition exhibits a pronounced hysteresis. Slower shear rate sweep leads to a reduced hysteresis loop, as expected for subcritical transitions~\cite{Rao90}. Another strong indication of the subcritical character of the transition is given in Fig.~\ref{fig2}b where, for imposed Weissenberg numbers in the transition range, the system exhibits bi-stability. As observed for polymer solutions, the transition towards the apparent disordered state is characterized by a sudden stress jump. The visual impression of spatiotemporal disorder in Fig.~\ref{fig1} inset-d is confirmed by a more careful analysis. Fig.~\ref{fig3}(a) shows average Fourier spectra of the intensity profiles along the vorticity direction, which exhibit power law decay over two decades in the wavevector domain. The power spectrum in wavevector seems to follow a power law up to $k_d\sim 100$~mm$^{-1}$. This suggests that for lengths below 10 $\mu$m, the dissipation process is different, leading to a faster relaxation of the fluctuations. 
In the case of elastic turbulence, $k_d$ is expected to be linked to the stress diffusion coefficient $\D$ by $k_d^{-1}\sim  \sqrt{\D\tau_R}$~\cite{Fouxon03}. When constitutive models are used to describe the dynamics of surfactant solutions, $\D$ is typically the coefficient of non-local diffusive terms~\cite{Fielding07}. Hence from the value of $k_d$ we observe here in the turbulent regime, we can compute a stress correlation length of the order of 10 $\mu$m. This estimation agrees with the stress correlation length (4 to 8 $\mu$m) obtained from completely different experiments in Ref.~\cite{Lerouge08,Ballesta2007,Masselon08}. Similarly, the intensity reflected by the sample at one point over time gives us some information about temporal disorder in the system. Fig.~\ref{fig3}(b) shows average Fourier spectra of the intensity at different points in the gap over time. The spectra exhibit power law decay over two decades in the frequency domain.
Furthermore, in order to get information on the fluctuations at the global scale, we have studied the time series of the macroscopic shear rate measured by the rheometer when imposing constant stresses. The inset of Fig.~\ref{fig4} gives the evolution of the amplitude of the shear rate fluctuations as a function of the shear stress. Below the transition to turbulence, fluctuations are small ($\Delta Wi$=1), less than 3$\%$ of the average value. In the transition range fluctuations are very high ($\Delta Wi$=11), up to 30$\%$ of the average value. In the fully turbulent regime, $i.e.$ $\sigma/G_0 >1$, fluctuations are as high as 15$\%$ of the average value ($\Delta Wi$=5). Average Fourier spectra of the fluctuations in this regime exhibit power law decay over almost two decades in the frequency domain, with an exponent $\beta$ of $-3.5$. This value of the exponent satisfies the criterion derived by Fouxon \textit{et al.} ($\beta<-3$) who have rationalized the spectra of turbulence in polymer solutions~\cite{Fouxon03}. \\
Hence, all the features highlighted for $Wi>Wi_c$ indicate that the dynamics in this regime are likely due to the same mechanisms driving the elastic turbulence in polymer solutions. Note however that the transitional pathway between the (1D) laminar and the turbulent regime in polymer solutions appears  different. In particular, when decreasing $Wi$, we do not observe solitary vortex pairs~\cite{Groisman97}, the relaxation of turbulence being  homogeneous. Differences might arise from the existence, for micelles, of two distinct relaxation mechanisms underneath $\tau_R$ ($\tau_r$ and $\tau_b$).

Let us summarize the complete scenario we observe~: in the shear banding regime, the flow is unstable with respect to perturbations along the vorticity direction and Taylor-like vortices develop in the high shear rate band. The stability of the flow seems to be recovered for $Wi_h<Wi<Wi_c$, namely when the high shear rate phase fully fills the gap, up to a critical Weissenberg number $Wi_c$ above which the system undergoes a transition towards a turbulent state. Taking into account the classical succession of instabilites expected with increasing Wi~\cite{Larson00,Morozov07}, from a purely 1D flow to a non trivial coherent flow and finally to a turbulent flow, the stability of the high shear rate phase in the range $Wi_h<Wi<Wi_c$ is intriguing. One explanation could be linked with the fact that instability thresholds strongly depend on boundary conditions. From the point of view of the high shear rate phase, the boundary conditions change with increasing $Wi$. Indeed, during the shear banding regime, the high shear rate phase is confined between a rigid wall (the inner rotating cylinder) and the viscous band that acts as a ``soft boundary". In contrast, for $Wi\gtrsim Wi_h$, the high shear rate phase is in direct contact with the two rigid walls of the Couette cell. In the same manner as many other instability examples, we can expect the instability threshold to be lower for soft boundaries than for rigid boundaries~\cite{Busse81}. A more precise picture following this trend could be given by an analysis focused on the high shear rate phase and considering the viscous phase as an elastic boundary~\cite{Kumar03}. Strictly, the intermediate stable regime may also suggest that instability in the plateau regime is due to the interfacial mechanism, and that the bulk instability of the high shear rate branch for $Wi>Wi_c$ is a classical elastic turbulence. The stability analyses that uncovered the interfacial mechanism have been performed in planar geometries. Very recently S. Fielding extended her original calculation to Taylor-Couette flow to test the effect of curvature on the interfacial mechanism~\cite{Fielding2010}. Using a unique framework, she  suggests that both interfacial and bulk elastic mechanisms could be observed for different values of streamlines curvature and normal stresses. This first theoretical work dealing with shear banding and elastic instabilities together predicts that in a curved geometry similar to ours, the interfacial mechanism is less favorable. In conclusion, our findings establish clearly that elastic instabilities and shear banding are a lot more intertwined than presupposed. These findings, together with the most recent theoretical studies, now rationalize 3D disturbances during shear banding in flows with curved streamlines as being driven by elastic instability.

\textbf{\normalsize{Acknowledgments }} \\
The authors thanks J.L. Counord for the building of the optical device, B. Abou and M. Receveur for the fast camera, S. Asnacios, S. Fielding and A.N. Morozov for fruitful discussions, and the ANR JCJC-0020 for financial support.

\end{document}